\documentclass[twocolumn,showpacs,aps]{revtex4}
\usepackage{graphicx}
\usepackage{bm}
\usepackage{amsmath}
\usepackage{amsfonts}
\usepackage{amssymb}
\usepackage[utf8]{inputenc}
\usepackage[american]{babel}
\usepackage[dvips]{color}

\newcommand{\mbf}[1]{\mathbf{#1}}
\newcommand{\ket}[1]{|#1\rangle}

\newcommand{\braket}[3]{\langle#1|#2|#3\rangle}
\newcommand{\scalar}[2]{\langle#1|#2\rangle}

\newcommand{\op}[1]{|#1\rangle\langle#1|}
\newcommand{\opp}[2]{|#1\rangle\langle#2|}

\newcommand{\tr}[2]{\mbox{trace}_{#1}\left(#2\right)}

\begin{document}

\title{Asymptotic entanglement in a two-dimensional quantum walk}
\author{M. Annabestani}
\email{annabestani@modares.ac.ir}
\author{M.R. Abolhasani}
\affiliation{Dept. of Physics, Basic Sciences Faculty, Tarbiat Modarres University, Tehran, Iran}
\author{G. Abal}\email{abal@fing.edu.uy}
\affiliation{Instituto de Física, Universidad de la República, Montevideo, Uruguay}
\date{\today}

\begin{abstract}
The evolution operator of a discrete-time quantum walk involves a conditional shift in position space which entangles the ``coin'' and position degrees of freedom of the walker. After several steps, the coin-position entanglement (CPE) converges to a well defined value which depends on the initial state. In this work we provide an analytical method which allows for the exact calculation of the asymptotic reduced density operator and the corresponding CPE for a discrete-time quantum walk on a two-dimensional lattice. We use the von Neumann entropy of the reduced density operator as an entanglement measure. The method is applied to the case of a Hadamard walk for which the dependence of the resulting CPE on initial conditions is obtained. Initial states leading to maximum or minimum CPE are identified and the relation between the coin or position entanglement present in the initial state of the walker and the final level of CPE is discussed. The CPE obtained from separable initial states satisfies an additivity property in terms of CPE of the corresponding one-dimensional cases. Non-local initial conditions are also considered and we find that the extreme case of an initial uniform position distribution leads to the largest CPE variation. 
\end{abstract}



\maketitle

\section{Introduction}
\label{sec:intro}

The quantum walk (QW) is a reversible process usually introduced as a quantum analog of a Markovian process \cite{Kempe03}. Several QW-based algorithms for meaningful problems have been developed \cite{Shenvi03,Ambainis03,AKR,Ambainis05,Childs04,Tulsi08} and they perform better than the best classical alternatives. In some cases, exponential speedups may be obtained \cite{Kempe02,Childs02,Childs03}. There are two versions of QW, based on discrete time \cite{Nayak} or continuous time \cite{FG98}. Both have similar dynamical properties and the later can be obtained by a suitable limiting process from the discrete-time walk \cite{Strauch,Childs08}. A second level of classification is based on the network over which the walk takes place. The QW on a linear chain is the simplest possible configuration \cite{Nayak,qw-markov,Konno,Venegas}, but other topologies such as cycles \cite{cycle1,cycle2}, two-dimensional lattices \cite{Mackay,Carneiro,Omar,Amanda,Konno2}, or $n$-dimensional hypercubes \cite{Moore,Marquezino} have also been investigated.

Certain aspects of the QW can be simulated using optical analogies \cite{KRS03,Knight03}, but to capture the true quantum nature of a QW one must deal with entanglement, a key resource for quantum information processing. The discrete time versions of a QW require an auxiliary ``coin" subspace ${\cal H}_C$. The evolution in the position subspace, ${\cal H}_P$, consists of a conditional shift determined by the amplitudes of the states in ${\cal H}_C$. The Hilbert space of the system is ${\cal H}={\cal H}_C\otimes {\cal H}_P$. Many QW applications involve partial measurements or noise events which selectively affect a part of the system (frequently, either the coin or position subspaces). The way the system is affected depends on the degree of entanglement between coin and position just before the event. A precise knowledge of this entanglement is required to gain control over the long-time behavior of the system, including the way it responds to partial measurements or selective noise events. It has also been recently suggested that the QW protocol may be useful as entanglement generator in two-body \cite{VB09} or even in many-body systems \cite{Goyal}. 

In a quantum walk the conditional shift operation generates entanglement between the coin and position (CPE) degrees of freedom of the walker. After several steps, it converges to a well defined value which, for a given evolution operator, is determined by the initial state ~\cite{Carneiro}. For a one-dimensional (Hadamard) QW, the dependence of CPE on the initial state has been characterized using analytical methods ~\cite{Abal06-ent}. However, most algorithmic applications require higher dimensions. Furthermore, higher dimensional QW's open new possibilities, such as the preparation of initially entangled states within the coin (CCE) or position subspaces (PPE). 

This paper deals with the characterization of long time CPE in a two-dimensional discrete-time QW, such as the one used in recent algorithmic proposals ~\cite{AKR,Tulsi08}. This system describes the motion of a quantum walker on a two-dimensional (2D) lattice or, alternatively, of two independent walkers along linear lattices. Our approach provides exact results for the two-dimensional QW with arbitrary initial conditions and coin operations. We provide several examples for a two-dimensional Hadamard quantum walk.

Any real implementation of a quantum system must deal with the issue of decoherence, which tends to destroy quantum correlations. The entanglement decay due to noise has been numerically investigated in one-dimensional systems \cite{MK06} and there are many papers on the important subject of decoherent quantum walks, see Ref.~\cite{Kendon-review} for a recent review. In this work, we restrict our attention to a coherent quantum walk on two dimensions and obtain the dependence of the asymptotic CPE on the initial coin state for local and non-local initial positions. In particular, we show that the non-local case can be easily obtained from the local case by adding a weight factor to the final integration.

This work is organized as follows. Section II defines the two-dimensional quantum walk and provides the required formalism in Fourier space leading to the long-time reduced density operator for arbitrary coin operations and initial states. In Section III, the method is applied to obtain the asymptotic CPE entanglement of a two-dimensional Hadamard walk as function of several families of initial states. This section includes a discussion on the additivity of CPE for separable cases. Finally, in Section IV, we summarize our results and present our conclusions.

\section{Asymptotic entanglement}
\label{sec:formalism}
The two-dimensional quantum walk is defined in terms of a discrete lattice whose sites are labeled by pairs of integers $(x,y)$. One can think in terms of two particles moving along two lines or in terms of a single particle moving on a plane. For definiteness, in this work we adopt the language of a single particle moving on a two-dimensional lattice. The set of orthonormal states $\{\ket{x,y}\}$ spans the position subspace, ${\cal H}_P$, of the walker. The ``coin'' degree of freedom is represented by a two-qubit space, ${\cal H}_C$, spanned by four orthonormal states which we label as $\{\ket{L,L},\ket{L,R},\ket{R,L},\ket{R,R}\}$. This nomenclature is motivated by the quantum walk on a line, where $\ket{L}$ and $\ket{R}$ are associated with left or right displacements respectively. 

The Hilbert space for the system is ${\cal H}={\cal H}_P\otimes{\cal H}_C$. A generic state is
\begin{equation}\label{eq:gen_state}
 \ket{\Psi}=\sum_{x,y}\sum_{j=1}^4 f_j(x,y)\, \ket{x,y}\otimes\ket{j},
\end{equation}
where the first sum runs over all lattice sites and we used a compact notation ($\ket{j},\; j=1\ldots 4$) for the coin states $\{\ket{L,L},\ket{L,R},\ket{R,L},\ket{R,R}\}$. One step of the evolution is described by
\begin{equation}\label{eq:evol_generic}
 \ket{\Psi(n+1)}=U\ket{\Psi(n)},
\end{equation}
where $n$ is a step counter and the evolution operator is 
\begin{equation}\label{eq:Uop}
 U=S\cdot (I_P\otimes U_C)
\end{equation} 
with $I_P$ the identity operator in ${\cal H}_P$. The evolution combines a unitary coin operation $U_C$ in ${\cal H}_C$ with a shift operator
\begin{equation}\label{eq:shift}
 \begin{split}
  S &= \sum_{x,y}\left\{\opp{x-1,y}{x,y}\otimes\op{1}+\opp{x,y+1}{x,y}\otimes\op{2}\right.\\
&\left.+\opp{x,y-1}{x,y}\otimes\op{3}+\opp{x+1,y}{x,y}\otimes\op{4}\right\}
 \end{split}\raisetag{12pt}
\end{equation}
which performs the conditional displacements determined by the coin state. The correspondence between coin states and displacements is not unique. The shift operator defined in eq.~(\ref{eq:shift}) represents the extension of a one-dimensional walk to two dimensions in a $45^o$ rotated lattice. This choice for $S$ simplifies the description in the Fourier representation, without loss of generality. 

Each spatial component of the wavevector,  $\ket{\psi_{x,y}}\equiv\scalar{x,y}{\Psi}=\sum_{j=1}^4 f_j(x,y)\ket{j}$, evolves as
\begin{eqnarray}\label{1step-r}
 \ket{\psi_{x,y}(n+1)}&=&M_1\,\ket{\psi_{x+1,y}(n)}+M_2\,\ket{\psi_{x,y-1}(n)}\\
&&~ +\,M_3\,\ket{\psi_{x,y+1}(n)}+M_4\,\ket{\psi_{x-1,y}(n)}.\nonumber
\end{eqnarray}
in terms of four operators $(M_i)$ acting in ${\cal H}_C$. For a generic coin operation, $U_C=\sum_{i,j}C_{i,j}\opp{i}{j}$, these operators are
\begin{equation}\label{Mi}
M_i\equiv \op{i}\,U_C=\sum_{j=1}^4 C_{i,j} \opp{i}{j}.
\end{equation}

The Fourier transform, as first noted in this context by Nayak and Vishwanath \cite{Nayak}, is extremely useful when single-step displacements are involved because the evolution operator is diagonal in $k$-space. The Fourier transform of the position eigenstates is
\begin{equation}\label{eq:k_states}
 \ket{\mbf{k}}\equiv\ket{k_x,k_y}=\sum_r  e^{i\mathbf{k}\cdot \mbf{r}}\ket{\mbf{r}}
\end{equation}
where $\mbf{r}$ is the vector with integer components $(x,y)$ and $\mathbf{k}$ is a vector with real components $(k_x,k_y)$ in the interval $[-\pi,\pi]$. The $k^{th}$ component of the wavevector, eq.~(\ref{eq:gen_state}), is
\begin{equation}
\ket{\psi_k}\equiv\sum_{j=1}^4 \tilde f_j(\mbf{k})\,\ket{j}
\end{equation}
has amplitudes $\tilde f_j(\mbf{k})\equiv\sum_r e^{-i\mbf{k}\cdot\mbf{r}} f_j(\mbf{r})$. As mentioned before, the linear map in  eq.~(\ref{1step-r}) can be expressed as the action of a diagonal operator in k-space
\begin{eqnarray}\label{1step-k}
 \ket{\psi_k(n+1)}&=&U_k\ket{\psi_k(n)}\nonumber\\
&=&\left( e^{-ik_x}M_1+e^{ik_y}M_2\right.\\
&&\qquad \left. +\,e^{-ik_y}M_3 +e^{ik_x}M_4\right)\ket{\psi_k(n)}.\nonumber
\end{eqnarray}
This equation defines a unitary operator $U_k$ which is represented by a $4\times 4$ matrix. The basic idea behind our approach is to use the spectral decomposition of $U_k$ to obtain information about the long-time evolution of the system. 

Let us consider the eigenproblem for the unitary operator $U_k$ with eigenvalues $e^{i\omega_k}$ and corresponding normalized eigenvectors
\begin{equation}\label{eq:egv}
 \ket{e(\omega_k)}=\sum_{j=1}^4 \alpha_j(\omega_k)\,\ket{j}.
\end{equation}
Using the spectral decomposition for $U_k$, the state of the system after $n$ steps is
\begin{equation}\label{k-evol}
\ket{\psi_k(n)}= U_k^n\ket{\psi_k(0)}=\sum_{\{\omega_k \}} e^{i\omega_k n} F(\omega_k)\;\ket{e(\omega_k)},
\end{equation}
where the sum is over the set of eigenvalues of $U_k$ and
\begin{equation}\label{eq:F}
F(\omega_k)\equiv\scalar{e(\omega_k)}{\psi_k(0)}=\sum_{j=1}^4 \alpha_j^*(\omega_k)\,\tilde f_j(\mbf{k}).
\end{equation}
Note that $\tilde f_j(\mbf{k})$ are the Fourier--transformed initial amplitudes, so $F(\omega_k)$ contains all the information about the initial state.


Since we deal with pure states only, we use the von Neumann entropy of the reduced density operator, or entropy of entanglement, defined as
\begin{equation}\label{eq:ent}
 E\equiv -\tr{}{\rho_c\log_2\rho_c}
\end{equation}
as a measure of coin-position entanglement (CPE). In this expression, $\rho_c=\tr{P}{\rho}$ is the reduced density operator obtained from $\rho=U^n\rho_0 U^\dagger\,^n$ by tracing out the position degrees of freedom. Since $\rho_c$ has dimension 4, this quantity is $E\in[0,2]$, i.e. $E=0$ for a product state and $E=2$, for a maximally entangled state. For a quantum walk on a line, $\rho_c$ is two-dimensional and $\protect{E\in [0,1]}$ and for $\rho_c$ with dimension $d$, $E$ is in the interval  $[0,\log_2 d]$. 

The inverse Fourier transform required to generate $\rho$ at arbitrary times can not be computed exactly, but since our main interest is to obtain the entropy of entanglement, we can avoid this issue by using Parseval's theorem. The reduced density operator, $\rho_c$, is obtained by taking the trace in $k$-space of the density operator
\begin{equation}\label{eq:rho_c1}
\rho_c=\tr{K}{\rho}=\int\frac{d^2\mbf{k}}{4\pi^2}\, \op{\psi_k},
\end{equation}
where $\tr{K}{\cdot}$ traces over $(k_x,k_y)$ and the integration in $\protect{d^2\mathbf{k}=dk_y\,dk_x}$ has limits $[-\pi,\pi]$. This expression can be evaluated after many steps of the evolution in a form completely analogous to the one used in Ref.~\cite{Abal06-ent} for a quantum walk on the line. From eq.~(\ref{k-evol}), after $n$ steps,
\begin{equation*}\label{op-psik}
\op{\psi_k}=\sum_{\{\omega_k,\,\omega_k^\prime\}} e^{i(\omega_k - \omega_k^\prime)n}\; F(\omega_k) F^*(\omega_k^\prime)\; \opp{e(\omega_k)}{e(\omega_k^\prime)}.
\end{equation*}
In the asymptotic limit $n\gg 1$, according to the stationary  phase theorem, only terms with $\omega_k=\omega_k^\prime$ contribute in eq.~(\ref{eq:rho_c1}) as discussed in detail in Ref.~\cite{Nayak}. Thus
\begin{equation}\label{eq:rho_c}
 \hat\rho_c=\int\frac{d^2\mathbf{k}}{4\pi^2}\, \sum_{\{\omega_k\}} \left\vert F(\omega_k)\right\vert ^2\op{e(\omega_k)},
\end{equation}
where we use a caret $(\hat~)$ to indicate that the asymptotic limit $\hat\rho_c\equiv\lim_{n\rightarrow\infty} \rho_c(n)$ has been taken. According to eq.~(\ref{eq:egv}), the matrix elements of $\hat\rho_c$ are 
\begin{equation}\label{eq:hat_rho_C}
\braket{i}{\hat\rho_c}{j}=\int\frac{d^2\mathbf{k}}{4\pi^2}\,{\cal P}_{i,j}(\mathbf{k})
\end{equation}
in terms of 
\begin{equation}\label{eq:Pij}
{\cal P}_{i,j}(\mathbf{k})\equiv
\sum_{\{\omega_k\}}\left\vert F(\omega_k)\right\vert^2\, \alpha_i(\omega_k)\alpha^*_j(\omega_k). 
\end{equation}
Note that these expressions satisfy ${\cal P}_{i,j}={\cal P}_{i,j}^*$, as required by the hermiticity of $\hat\rho_c$. 

At this point we need to specify the form of the initial state. Since the main interest of this work is to characterize the long-time coin-position entanglemet (CPE) generated by the evolution of the quantum walk, we shall consider only separable coin-position initial states, i.e.
\begin{equation}\label{in-state}
 \ket{\Psi(0)}=\ket{\psi}\otimes\ket{\chi}
\end{equation}
with initial position $\ket{\psi}=\sum_{\mathbf{r}} a(\mathbf{r})\ket{\mathbf{r}}$ and initial coin $\ket{\chi}=\sum_j c_j\ket{j}$. These states have no CPE and $\protect{f_j(\mathbf{r})=c_j\, a(\mathbf{r})}$ in eq.~(\ref{eq:gen_state}), so that their $k$-component is
\begin{equation}\label{eq:krep-CPseparable}
 \ket{\psi_k(0)}=\tilde a(\mathbf{k})\;\ket{\chi}
\end{equation}
in terms of the Fourier-transformed initial amplitudes $\tilde a(\mathbf{k})\equiv\sum_\mathbf{r} e^{-i\mathbf{k}\cdot \mathbf{r}} a(\mathbf{r})$. From eq.~(\ref{eq:F}) one readily obtains for this kind of initial conditions,
\begin{equation}\label{eq:F-nolocal}
|F(\omega_k)|^2=|\tilde a(\mathbf{k})|^2\sum_{j,l}\alpha_j^*(\omega_k)\alpha_l(\omega_k)\, c_jc_l^*.
\end{equation}
From this expression for a given coin operation and initial state one can obtain $\hat\rho_c$ and, after diagonalization, the entropy of entanglement $E$, from eq.~(\ref{eq:ent}). Thus, the dependence on the initial conditions of the CPE generated by the evolution after many steps can be explored using this method. We emphasize that it can be generalized in a straightforward form to arbitrary dimensions and coin operations $U_C$, as long as one is able to solve the eigenproblem for $U_k$ analytically.

\section{Dependence on the initial state}
\label{sec:istate}

In order to discuss the dependence of CPE on the initial state, we must specify a coin operation. 
We shall consider the case of a two-dimensional Hadamard walk. When applied to the basis states of a single qubit, a Hadamard operation generates the balanced superpositions $\protect{H\ket{L}=(\ket{L}+\ket{R})/\sqrt{2}}$ and $\protect{H\ket{R}=(\ket{L}-\ket{R})/\sqrt{2}}$. This coin operation is a common choice in the literature of one-dimensional quantum walks \cite{Nayak, Konno-sym, qw-markov} and information on the asymptotic coin-position entanglement (CPE) level is available for this case \cite{Carneiro, Abal06-ent}. Our method is also applicable to other cases of interest, such as the Grover or DFT coin operations \cite{Carneiro}. 

For a quantum walk in two spatial dimensions the coin operation $U_C=H\otimes H$ is a natural extension of the Hadamard walk on a line. In this Section, we apply the general formalism previously outlined in the previous section to this case and characterize the asymptotic CPE level for several initial conditions.

From eqs.~(\ref{Mi}) and (\ref{1step-k}), the explicit form for the operator $U_k$ for $U_C=H\otimes H$ is
\begin{equation}\label{eq:Uk}
U_k=\frac{1}{2}\left( \begin{array}{cc}
 \begin{array}{cc}
e^{-ik_x} & e^{-ik_x} \\
e^{ik_y} & -e^{ik_y} \end{array}
 &  \begin{array}{cc}
e^{-ik_x} & e^{-ik_x} \\
e^{ik_y} & {-e}^{ik_y}
\end{array}\\
 \begin{array}{cc}
e^{-ik_y} & e^{-ik_y} \\
e^{ik_x} & -e^{ik_x} \end{array}
 &  \begin{array}{cc}
{-e}^{-ik_y} & {-e}^{-ik_y} \\
-e^{ik_x} & e^{ik_x} \end{array}
 \end{array}
\right).
\end{equation}
The eigenproblem for this operator can be solved exactly. Its four eigenvalues are $\protect{\{ e^{i\omega_+}, e^{-i\omega_+}, e^{i\omega_-}, e^{-i\omega_-}\}}$ with
\begin{eqnarray}\label{eq:cos}
\cos\omega_\pm&=&\frac14\left(\cos k_x -\cos k_y\pm\sqrt{\Delta_k}\right),\\
\Delta_k&\equiv& \cos^2 k_x  + 6\,\cos k_x\cos k_y + \cos^2 k_y  + 8.\nonumber
\end{eqnarray}
As usual, we shall refer to the phase of any of these eigenvalues by the symbol $\omega_k$. The normalized eigenvectors are of the form (\ref{eq:egv}) with components
\begin{eqnarray}\label{eq:eigenvec}
\alpha_1&=&\frac{1}{N}\left[1-e^{2i\omega_k}\right]\\
\alpha_2&=&\frac{1}{N}\left[-1 +  e^{i\omega_k}\left(e^{ik_x}-e^{ik_y}\right) +  e^{2i\omega_k}\, e^{i(k_x+k_y)}\right]\nonumber\\
\alpha_3&=&\frac{1}{N}\left[-1 +  e^{i\omega_k}\left(e^{ik_x}-e^{-ik_y}\right) +  e^{2i\omega_k}\, e^{i(k_x-k_y)}\right]\nonumber\\
\alpha_4&=&\frac{1}{N}\left[1+e^{i\omega_k}\left(e^{ik_y}+e^{-ik_y}\right) + \right.\nonumber\\
&&~ \left. e^{2i\omega_k}\left(1-e^{i(k_x+k_y)}-e^{i(k_x-k_y)}\right) -2e^{3i\omega_k}\,e^{ik_x}         \right].\nonumber
\end{eqnarray}
The normalization constant $N$ is a positive real, chosen so that $\sum_{j=1}^4 |\alpha_j|^2=1$. 

\subsection{Separable initial states}

Let us first consider in the detail the simple case of an initial state localized at the origin $(x=y=0)$ with initial coin state $\ket{\chi}=\ket{L,L}$, or $c_1=1$ and $c_j=0$ for $j\ne 1$. For a localized state, $\tilde a(\mbf{k})=1$ and the projection on k-space, eq.~(\ref{in-state}), is simply $\protect{\ket{\psi_k}=\ket{\chi}}$. Thus in this case, eq.~(\ref{eq:F-nolocal}) reduces to
\begin{equation}\label{eq:Fex}
|F(\omega_k)|^2 = 2(1-\cos(2\omega_k)).
\end{equation}
This expression is used in eq.~(\ref{eq:Pij}) to find
\begin{eqnarray}\label{eq:P00}
\mathcal{P}_{1,1}(\mathbf{k})&\equiv&\sum\limits_{\omega_k}\left|{F\left(\omega_k\right)}\right|^2\cdot\left|\alpha_1(\omega_k)\right|^2 \\
&=& \Delta_k^{-1}\left(\cos^2 k_x + 4\cos k_y \cos k_x + \cos^2 k_y  + 3\right).\nonumber
\end{eqnarray}
After averaging over $\mathbf{k}$, the matrix element of the reduced density operator is obtained,
\begin{equation}\label{eq:C1}
\hat\rho_c(1,1)=\int \frac{d^2\mathbf{k}}{4\pi^2}\,\mathcal{P}_{1,1}(\mathbf{k})\equiv C_1=\frac{9-4\sqrt{2}}{8}\, .
\end{equation} 
The other independent elements of $\hat\rho_c$ may be calculated in the same form,
\begin{eqnarray}\label{eq:rhoC-0}
\hat\rho_c(1,2)=\hat\rho_c(1,3)\equiv C_2=\frac{5-3\sqrt{2}}{8}\nonumber\\
\hat\rho_c(1,4)=\hat\rho_c(2,3)\equiv C_3=\frac{3-2\sqrt{2}}{8}\nonumber\\
\hat\rho_c(2,2)=\hat\rho_c(3,3)\equiv C_4=\frac{2\sqrt{2}-1}{8}\\
\hat\rho_c(2,4)=\hat\rho_c(3,4)\equiv C_5=\frac{\sqrt{2}-1}{8}.\nonumber
\end{eqnarray}
The eigenvalues for $\hat\rho_c$ are $\lambda_1=1/2$, $\lambda_2=4C_3$ and $\lambda_3=\lambda_4=4C_5$, so that the asymptotic entropy of entanglement for this case is
\begin{equation*}\label{value-separable-Ent}
E= - \tr{}{\hat\rho_c \log_2 \hat\rho_c} \simeq 1.744
\end{equation*}
This quantity can be compared to the asymptotic entanglement of a one dimensional Hadamard walk that starts at the origin with $\ket{L}$ (or $\ket{R}$) as the initial coin. In this cases,  the asymptotic CPE is $E_0\simeq 0.872$, ~\cite{Abal06-ent, Carneiro}. Thus, for $\ket{L,L}$ we obtain exactly twice CPE as in the one-dimensional case. This is due to the fact that both the initial coin state and the coin operation $H\otimes H$ are separable and the quantum walk on the plane decomposes into two independent one-dimensional walks. In these cases, an additivity property applies for the asymptotic CPE as we discuss in more detail below.

Let us suppose that the initial state is separable with respect to both walkers, i.e.
\begin{equation}
 \ket{\Psi}=\ket{\Psi_1}\otimes\ket{\Psi_2}
\end{equation} 
where $\ket{\Psi_i}=\ket{\phi_i}\otimes\ket{\chi_i}$ for $i=1,2$. The states $\ket{\phi_i}$, spanned by $\{\ket{x}\}$, describe initial positions and the states $\ket{\chi_i}$, spanned by $\{\ket{L},\ket{R}\}$, are one-qubit initial coin states. If the coin operation can be written as a product two one-qubit operators,  $\protect{U_C=U_{C1}\otimes U_{C2}}$, the  evolution operator $U$ defined in eq.~(\ref{eq:Uop}), is separable, $U=U_x\otimes U_y$, with $U_x=S_x\cdot(I_x\otimes U_{C1})$ and a similar expression for $U_y$. Then, the separability of the initial state is preserved by the evolution. After $n$ steps, the state vector can be expressed as 
\begin{equation}
\ket{\Psi}=U_x^n\ket{\Psi_1(0)}\otimes U_y^n\ket{\Psi_2(0)},
\end{equation} 
and the reduced density operator 
\begin{equation}
\rho_c =\tr{P}{\op{\Psi}}=\rho_{C1}\otimes\rho_{C2},
\end{equation} 
is also separable, with $\rho_{C1}=\tr{x}{\op{\Psi_1}}$ and a similar expression for $\rho_{C2}$. The subadditivity property of the von Neumann entropy for a separable density operator \cite{N+C} implies, 
\begin{equation}\label{eq:addE}
 E(\rho_{C1}\otimes\rho_{C2})=E(\rho_{C1})+E(\rho_{C2}).
\end{equation} 
In other words, for a separable initial state and separable coin operation, the CPE for a two-dimensional QW can be obtained from the CPE of the underlying one-dimensional quantum walks.

One can use this property as a witness of correctness of our method for the two-dimensional case. Let us consider a a localized initial position with a generic separable initial coin state
\begin{equation}\label{eq:loc-sep}
 \ket{\Psi(0)}=\ket{0,0}\otimes\ket{\chi_1(\theta_1,\phi_1)}\otimes\ket{\chi_2(\theta_2,\phi_2)}
\end{equation} 
where $\protect{\ket{\chi_j}=\cos\theta_j\,\ket{L}+e^{i\phi_j}\sin\theta_j\,\ket{R}}$ for $j=1,2$ are generic one-qubit states.  The four real parameters that specify this initial state are restricted to the intervals $\protect{\theta_j\in[-\pi/2,\pi/2]}$ and $\phi_j\in\protect{[-\pi,\pi]}$.

According to eq.~(\ref{eq:addE}), the asymptotic CPE in this case can be computed from the CPE of a one dimensional quantum walk with initial condition $\protect{\ket{\Psi_1(0)}=\ket{0}\otimes\ket{\chi_1(\theta,\phi)}}$. As mentioned before, for this case an exact expression for the CPE has been obtained in Ref.~\cite{Abal06-ent}, in terms of the initial coin state and the quantity 
\begin{equation}\label{eq:r1}
 \lambda(\theta,\phi)=\frac12\left[1+\left(1-4(\Delta_0-2b_1^2 \sin 4\theta \cos\phi)\right)^\frac12\right].
\end{equation}  
The constants in this expression are $\Delta_0=(\sqrt{2}-1)/2$ and $b_1=(2-\sqrt{2})/4$ and the asymptotic CPE for this one-dimensional, localized case is given by \cite{Abal06-ent}
\begin{equation}\label{eq:E1}
 E_1(\theta,\phi)=-\lambda\log_2 \lambda - (1-\lambda)\log_2 (1-\lambda).
\end{equation} 

Using additivity, one can obtain the CPE for localized, separable initial conditions (eq.~\ref{eq:loc-sep}) from this expression as
\begin{equation}\label{eq:Eadd12}
 E(\theta_1,\phi_1;\theta_2,\phi_2)=E_1(\theta_1,\phi_1)+E_1(\theta_2,\phi_2).
\end{equation} 
The RHS of this equation can be evaluated using eqs.~(\ref{eq:E1}) and (\ref{eq:r1}) and the LHS can be evaluated as described in the last section. Thus, eq.~(\ref{eq:Eadd12}), which applies to separable localized initial conditions only, is a useful witness for the correctness of the proposed method. 

\begin{figure}[h]
\begin{center}
\includegraphics[width=8.5cm]{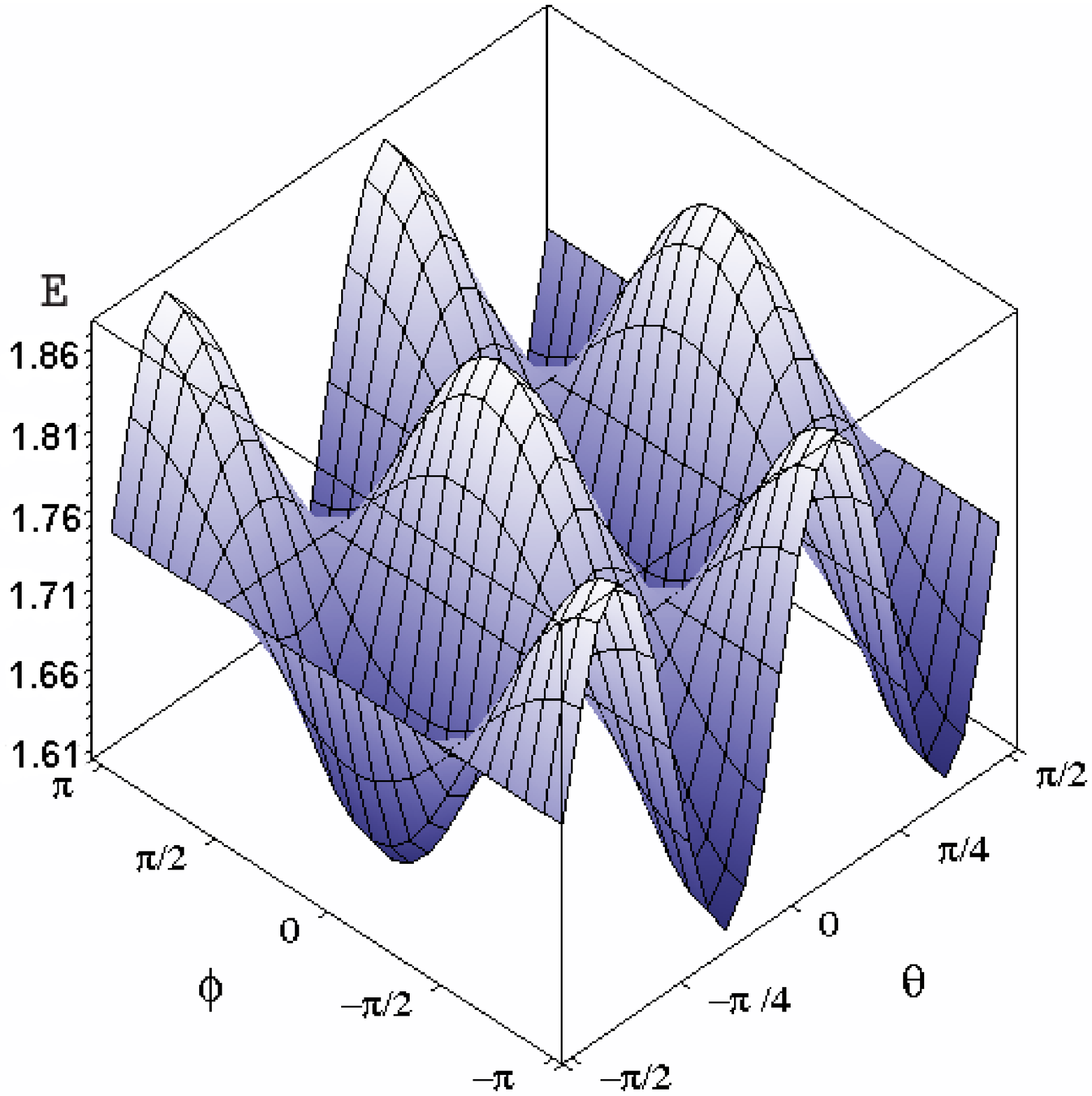}\\
\includegraphics[width=8.5cm]{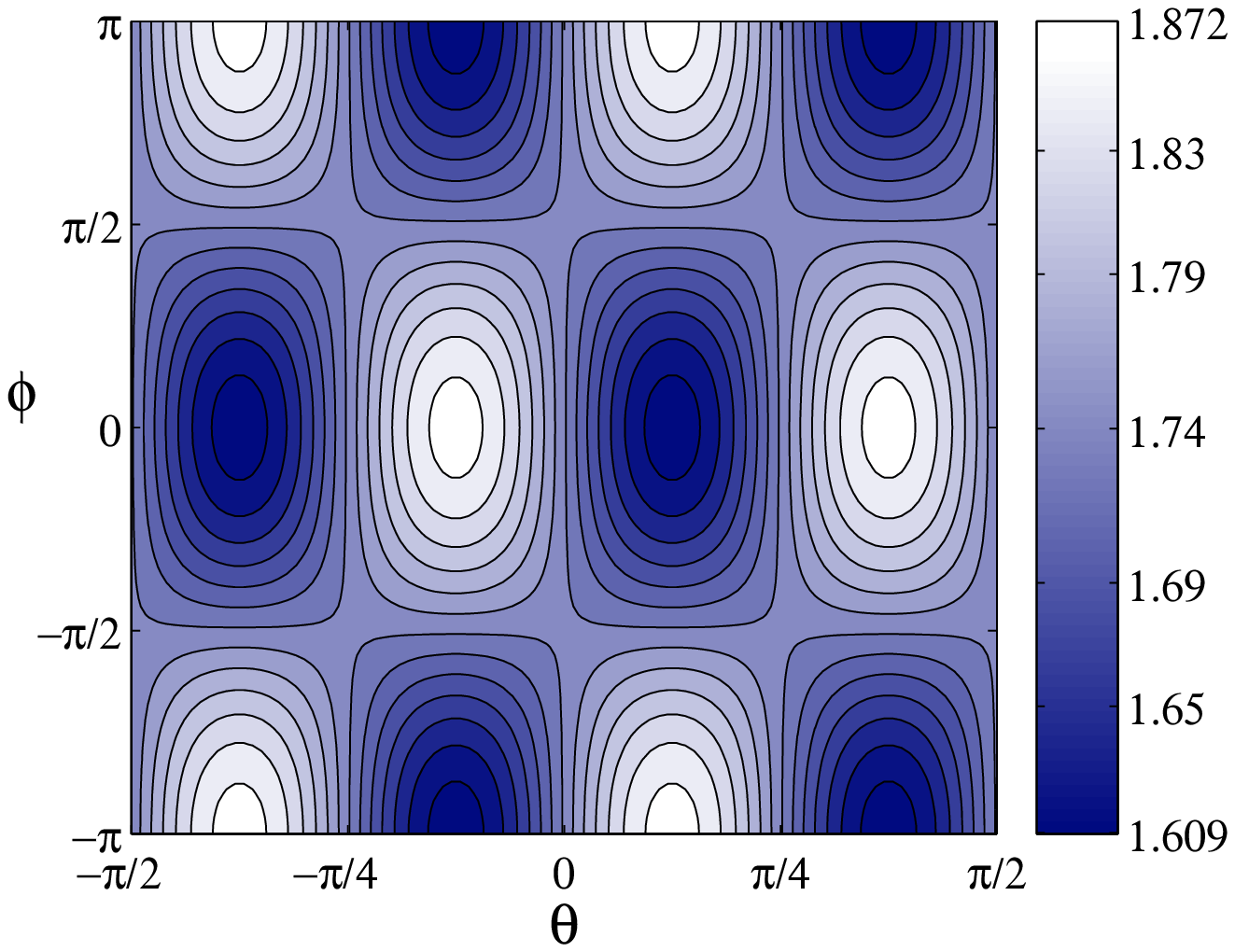}
\end{center}
\caption{\small (a) Asymptotic entropy of entanglement (CPE), $E(\theta,\phi)$, for a localized separable initial state, eq.~(\ref{eq:loc-sep}), with the particular initial coin state defined in eq.~(\ref{I}). (b)  Contour plot of the same surface.}
\label{FIG.L.Y}
\end{figure}

We shall now discuss some specific examples of the CPE dependence on the initial coin state. Let us consider a particular instance of eq.~(\ref{eq:loc-sep}) with $\theta_1=0$ and rename $(\theta_2,\phi_2)\rightarrow(\theta,\phi)$, so that the initial separable coin state is
\begin{equation}
\ket{\chi^{(I)}(\theta,\phi)}\equiv\ket{L}\otimes\left(\cos\theta\,\ket{L}+e^{i\phi}\sin\theta\,\ket{R}\right).
\label{I}
\end{equation}
For these states a simple calculation based on eq.~(\ref{eq:F-nolocal}) leads to
 \begin{equation*}\label{eq:F-I}
 |F|^2=|\alpha_1|^2\cos^2\theta + |\alpha_2|^2\sin^2\theta +
\sin(2\theta)\Re\left(\alpha_1^*\alpha_2 e^{-i\phi}\right).
 \end{equation*}
This expression is used in eq.~(\ref{eq:Pij}) to evaluate ${\cal P}_{ij}(\mathbf{k})$. After averaging in $\mathbf{k}$-space, the dependence of the asymptotic reduced density operator on the initial state parameters can be expressed in terms of the functions
\begin{eqnarray}\label{eq:hg}
h(\theta,\phi)&\equiv&\sin(2\theta)\cos\phi\nonumber\\
f(\theta,\phi)&\equiv&h(\theta,\phi)+\cos(2\theta) + 1 \\
g(\theta,\phi)&\equiv&f(\theta,\phi)-i\sqrt{2}\sin\phi\sin(2\theta)-1,\nonumber
\end{eqnarray}
and the constants $C_i$ defined in eqs.~(\ref{eq:C1})--(\ref{eq:rhoC-0}). The explicit form of the reduced density matrix is 
\begin{equation*}
 \hat\rho_c=\left(
\begin{array}{cccc}
 C_4+C_2 f &C_2 g  & C_5 +C_3 f &  C_3 g\\ 
 C_2 g^* & C_1-C_2 f & C_3 g^* & C_2-C_3 f \\ 
 C_5 +C_3 f & C_3 g & \frac18 +C_5 f & C_5 g \\ 
 C_3 g^* & C_2-C_3 f & C_5 g^* & C_4-C_5 f
\end{array}
\right).
\end{equation*} 
The exact diagonalization of this operator leads to the asymptotic entanglement as a function of the initial parameters, $E(\theta,\phi)$. Fig.~\ref{FIG.L.Y} shows the entropy of entanglement as a function of initial coin state defined in eq.~(\ref{I}). Alternatively, the surface shown in Fig.~\ref{FIG.L.Y} can also be calculated from the RHS of eq.~(\ref{eq:Eadd12}), 
\begin{equation}
 E(\theta,\phi)=E_0+E_1(\theta,\phi)
\end{equation} 
with $E_1$ given by eq.~(\ref{eq:E1}) and $\protect{E_0=E_1(0,\phi)\simeq 0.872}$.

\begin{figure}
\begin{center}
\includegraphics[width=8.5cm]{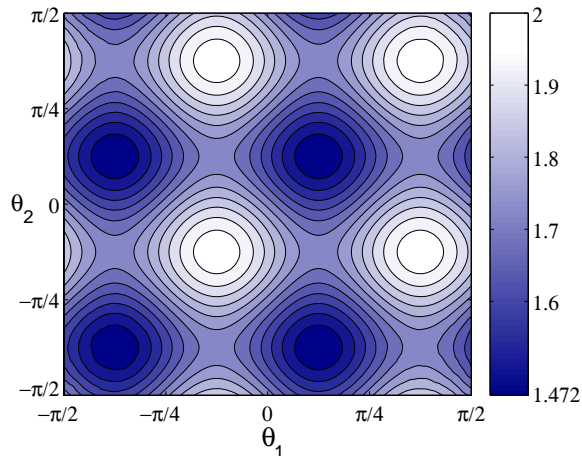}\\
\end{center}
\caption{\small Contour plot for the CPE, $E(\theta_1,\theta_2)$, obtained from the initial separable state eq.~(\ref{eq:loc-sep}), with $\phi_1=\phi_2=0$.}\label{fig:cont-sep}
\end{figure}

In Fig.~\ref{fig:cont-sep} we show the CPE for the case in which both initial coin states are superposition states. We have set $\phi_1=\phi_2=0$ in eq.~(\ref{eq:loc-sep}) and calculated $E(\theta_1,\theta_2)$ following the method outlined in the previous section. The result is consistent with the additivity property, eq.~(\ref{eq:Eadd12}). 

The previous argument, based on the separability of the motion, may be extended to any number $N\ge 2$ of quantum walkers with a separable initial state, $\ket{\Psi(0)}=\ket{\Psi_1(0)}\otimes\ldots\otimes\ket{\Psi_N(0)}$, evolving under a separable coin operation, $U_C=A_1\otimes\ldots\otimes A_N$. Then, the CPE can be obtained by addition of the corresponding one-dimensional CPE's, $E(\Psi(0))=\sum_{i=1}^N E_1(\Psi_i(0))$. The maximum entanglement of $E=N$ can be obtained of all initial states are properly prepared.

\subsection{Entangled initial states}
\begin{figure}
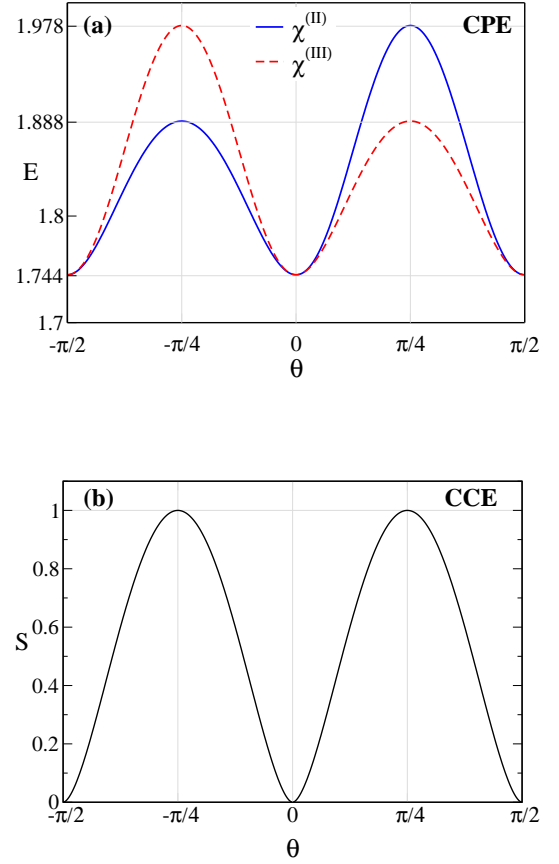

\begin{center}\vskip5mm
\includegraphics[width=7cm]{./fig/CPE-ent.eps}\\\vskip13mm
\includegraphics[width=7cm]{./fig/ee2q-new.eps}
\end{center}
\caption{\small (a) Asymptotic CPE, $E(\theta,\phi)$, for the families of initial coin states $\ket{\chi^{(II)}}$ (red, dashed line) and $\ket{\chi^{(III)}}$ (blue, full line).
(b) Initial CCE $S(\theta)$, defined in eq.~(\ref{eq:S}), for the same families of states.
}
\label{fig:compara-local}
\end{figure}

Up to this point we have discussed separable initial states only. Let us now discuss the effect on CPE of initially entangled states. Strict subadditivity of the von Neumann entropy holds only for separable density operators, so in this case CPE can not be obtained from one-dimensional calculations. Let us start with localized states 
\begin{equation}
 \ket{\Psi(0)}=\ket{00}\otimes\ket{\chi}
\end{equation} 
and consider two families of initial coin states,
\begin{eqnarray}
\ket{\chi^{(II)}}&\equiv&\cos\theta\,\ket{L,R}+e^{i\phi}\sin\theta\,\ket{R,L}\label{II}\\
\ket{\chi^{(III)}}&\equiv&\cos\theta\,\ket{L,L}+e^{i\phi}\sin\theta\,\ket{R,R}\label{III}
\end{eqnarray}
which describe entangled states in ${\cal H}_C$. Their entropy of entanglement is defined as
\begin{equation}\label{eq:S}
 S\equiv -\mbox{trace}_1\left[(\op{\chi})\log_2(\op{\chi})\right],
\end{equation} 
with the partial trace taken over either of the one-qubit coin subspaces spanned by $\{\ket{L},\ket{R}\}$. This quantity is normalized to one and measures the initial coin-coin entanglement (CCE) in $\ket{\chi}$. As shown in Fig.~\ref{fig:compara-local}~(b), it depends on the parameter $\theta$ alone. Note that, since the coin operation $U_C=H\otimes H$ is separable, the initial CCE is preserved by the evolution. In particular, the maximally entangled Bell states
\begin{eqnarray}
 \ket{\Psi^{\pm}}&\equiv& \frac{1}{\sqrt{2}}(\ket{L,R}\pm\ket{R,L})\label{eq:bell1}\\
\ket{\Phi^{\pm}}&\equiv& \frac{1}{\sqrt{2}}(\ket{L,L}\pm\ket{R,R}),\label{eq:bell2}
\end{eqnarray} 
included in these families, appear associated with maximum or minimum asymptotic CPE values.
\begin{figure}
\begin{center}
\includegraphics[width=8.5cm]{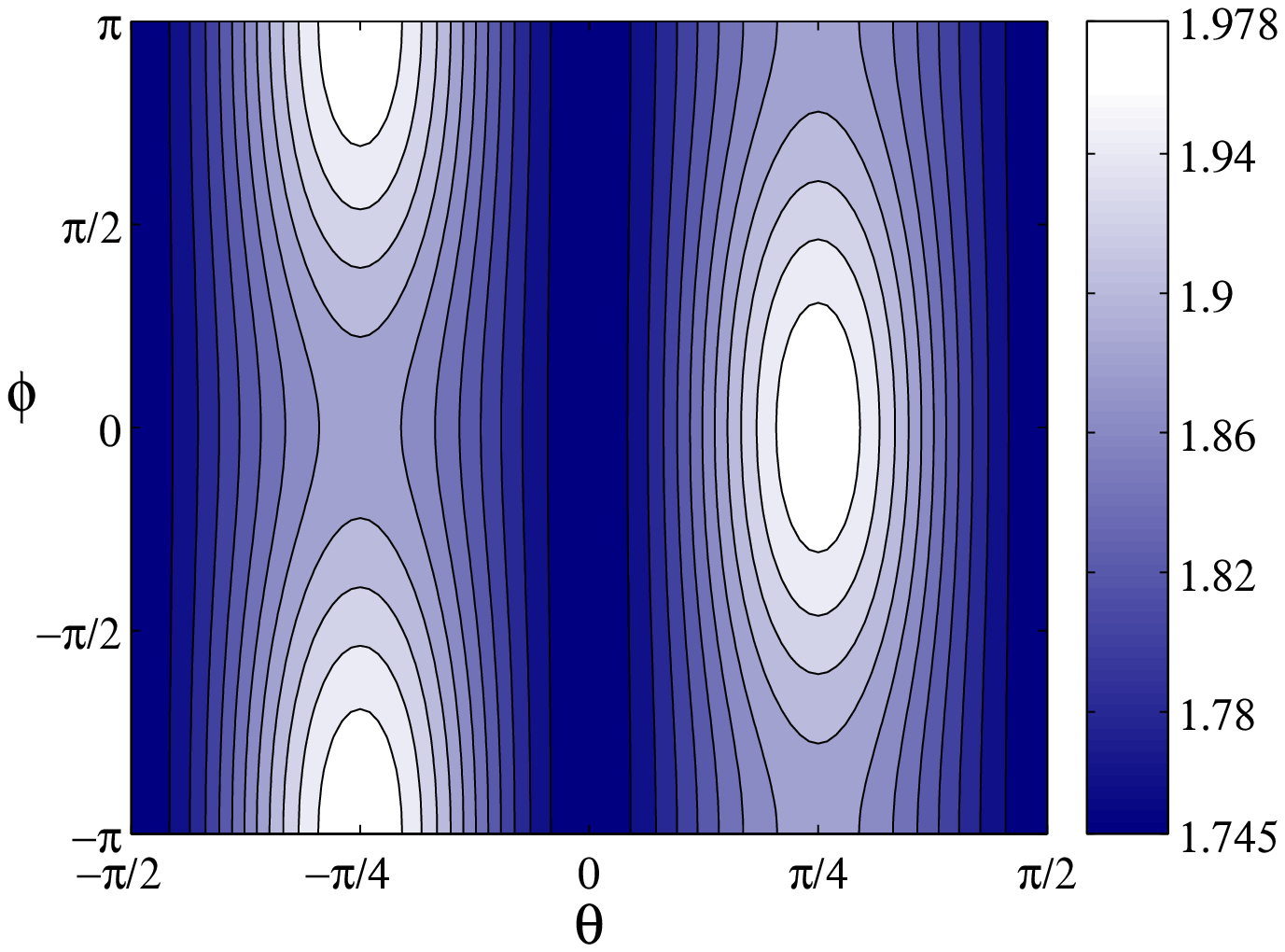}\\
\includegraphics[width=8.5cm]{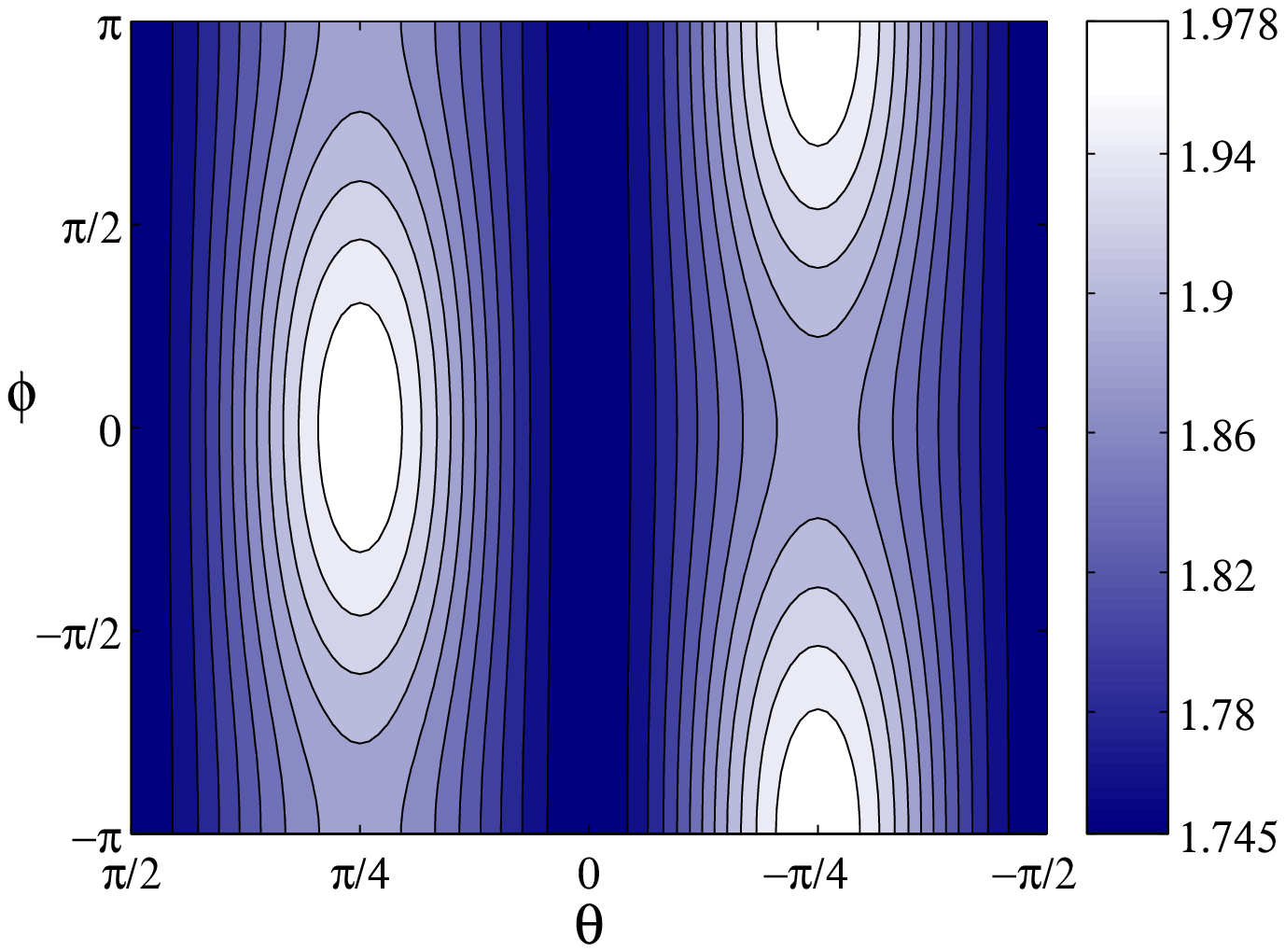}
\end{center}
\caption{\small Asymptotic entropy of entanglement resulting from entangled initial coin states. Upper panel: initial coin in family $\ket{\chi^{(II)}}$ defined in eq.~(\ref{II}). Lower panel: initial coin in family $\ket{\chi^{(III)}}$, defined in eq.~(\ref{III}).}
\label{FIG.L.PSI}
\end{figure}

For the first family, $\ket{\chi^{(II)}}$, straightforward evaluation of eq.~(\ref{eq:F-nolocal}) leads to
\begin{equation*}
 |F|^2=|\alpha_2|^2\cos^2 \theta + |\alpha_3|^2\sin^2 \theta +\sin(2\theta)\; \Re\left[\alpha_2^*\alpha_3\,e^{-i\phi}\right].
\end{equation*}
A similar expression holds for $\ket{\chi^{(III)}}$ with $(\alpha_2,\alpha_3)$ replaced by $(\alpha_1,\alpha_4)$. These expressions are used in eq.~(\ref{eq:Pij}) to obtain the elements of the hermitic matrix ${\cal P}(\mathbf{k})$. After the k-average is done, the long-time reduced density operator $\hat\rho_c$ is obtained. For $\ket{\chi^{(II)}}$, $\hat\rho_c$ has seven independent elements which can be expressed in terms of the constants $C_i$ and the functions $h,f,g$ defined in eqs.~(\ref{eq:hg}), as  
\begin{eqnarray}\label{rho-II}
\hat\rho_c(1,1)&=&\hat\rho_c(4,4)=C_4+C_3h\nonumber\\
\hat\rho_c(2,2)&=&C_3(f-2h)+\frac18 (f-h+1)\nonumber\\
\hat\rho_c(3,3)&=&-C_3f +\frac18 (h-f) +C_1\\
\hat\rho_c(1,2)&=&-\hat\rho_c^*(2,4)=C_3(g^*+1) -C_2(f-h) + C_5\nonumber\\
\hat\rho_c(1,3)&=&-\hat\rho_c^*(3,4)=C_3 (g+1) +C_5(f-h)- C_2\nonumber\\
\hat\rho_c(1,4)&=&-C_3\left(h+1\right)\nonumber\\
\hat\rho_c(2,3)&=&\left[3(h-1)+2(f-g^*)\right].\nonumber
\end{eqnarray}
Similar expressions can be obtained for the other entangled family, $\ket{\chi^{(III)}}$. After diagonalization of $\hat\rho_c$, an exact expression for the asymptotic CPE $E$ is obtained. For both cases, the dependence of this quantity on the initial coin state parameters $(\theta,\phi)$ is shown in Fig.~\ref{FIG.L.PSI}. It varies in the (approximate) range $[1.744, 1.978]$, with its minimum associated to separable initial coin states and its maximum associated to fully entangled initial coin states. Asymptotic CPE and initial CCE are related in the sense that maximum CPE is associated with maximum CCE while minimum CPE is associated with initial product states (i.e. no CCE). However, maximum initial CCE does not imply maximum asymptotic CPE; for the family $\chi^{(II)}$, maximum CPE is obtained from the Bell state $\ket{\Psi^+}$, but $\ket{\Psi^-}$, also fully entangled, leads to the intermediate CPE value $E\simeq 1.888$, see the upper panel Fig.~\ref{FIG.L.PSI}. For the symmetric family $\chi^{(III)}$, the same values are obtained but in this case $\ket{\Phi^-}$ yields maximum entanglement and $\ket{\Phi^+}$ leads to an intermediate value $E\simeq 1.888$, see the lower panel Fig.~\ref{FIG.L.PSI}.

A comparison between both panels in Fig.~\ref{FIG.L.PSI} shows that the relation  $E(\chi^{(III)};\theta,\phi)=E(\chi^{(II)};-\theta,\phi)$ is satisfied. From a mathematical point of view this may be traced to the fact that the reduced density operators $\rho_c(\chi^{(III)};\theta,\phi)$ and $\rho_c(\chi^{(II)};-\theta,\phi)$ have the same eigenvalues. We also note that both initial coins have the same CCE, as indicated by $S(\theta)$, which is an even function (see Fig.~\ref{fig:compara-local}~b). 

\begin{figure}
\begin{center}
\includegraphics[width=5.5cm]{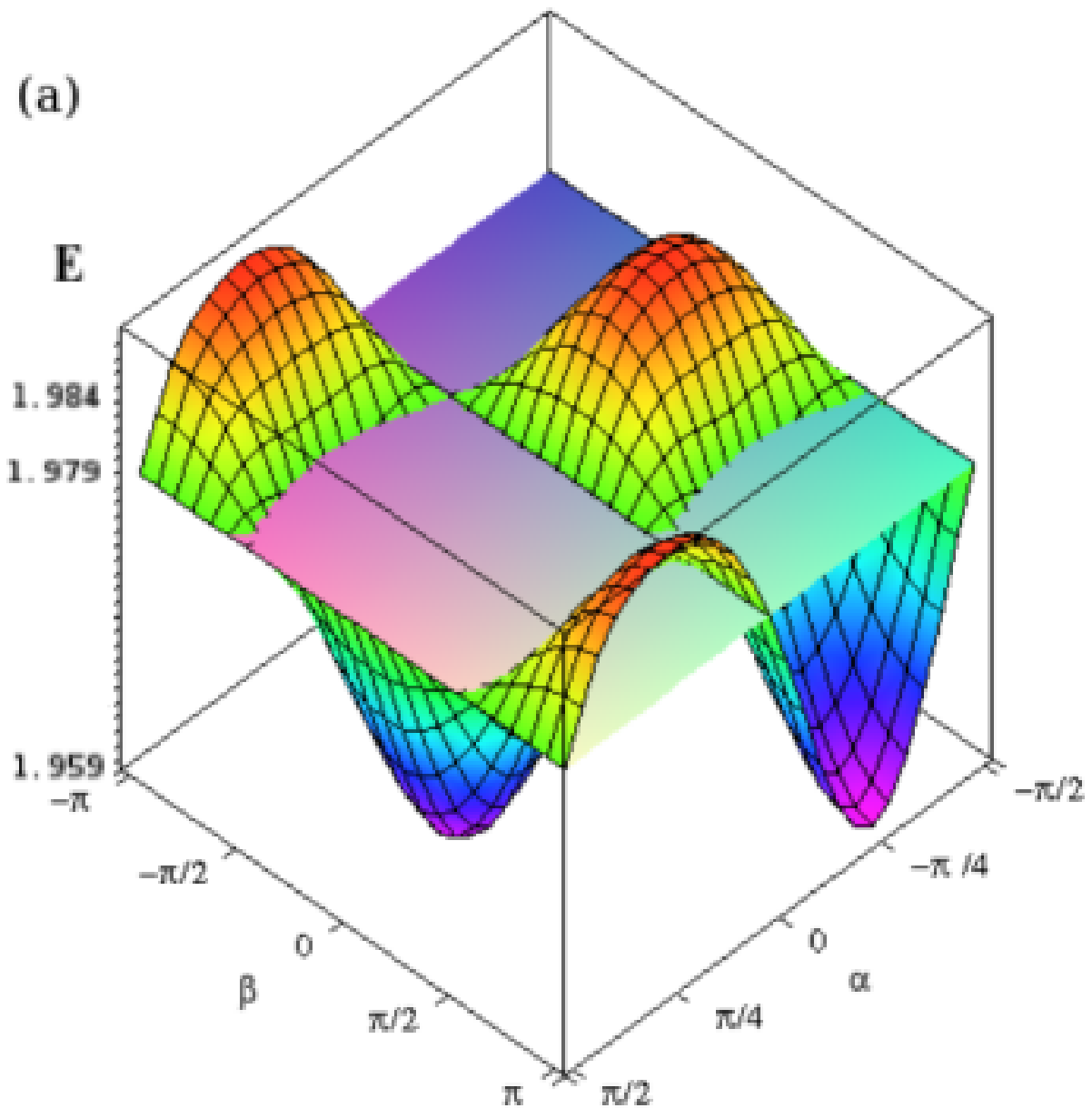}\\
\includegraphics[width=5.5cm]{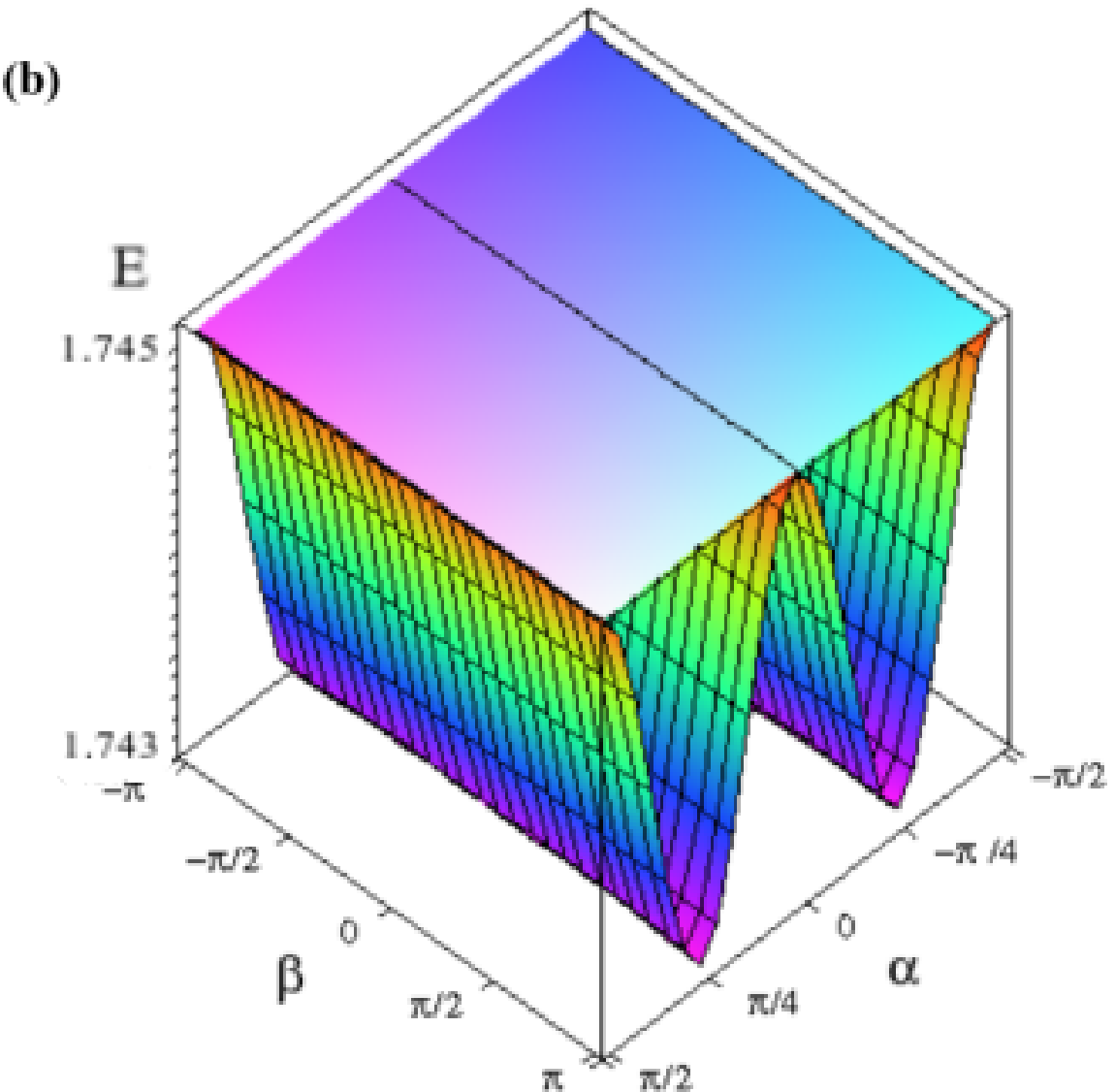}\\
\hskip3mm\includegraphics[width=5.8cm]{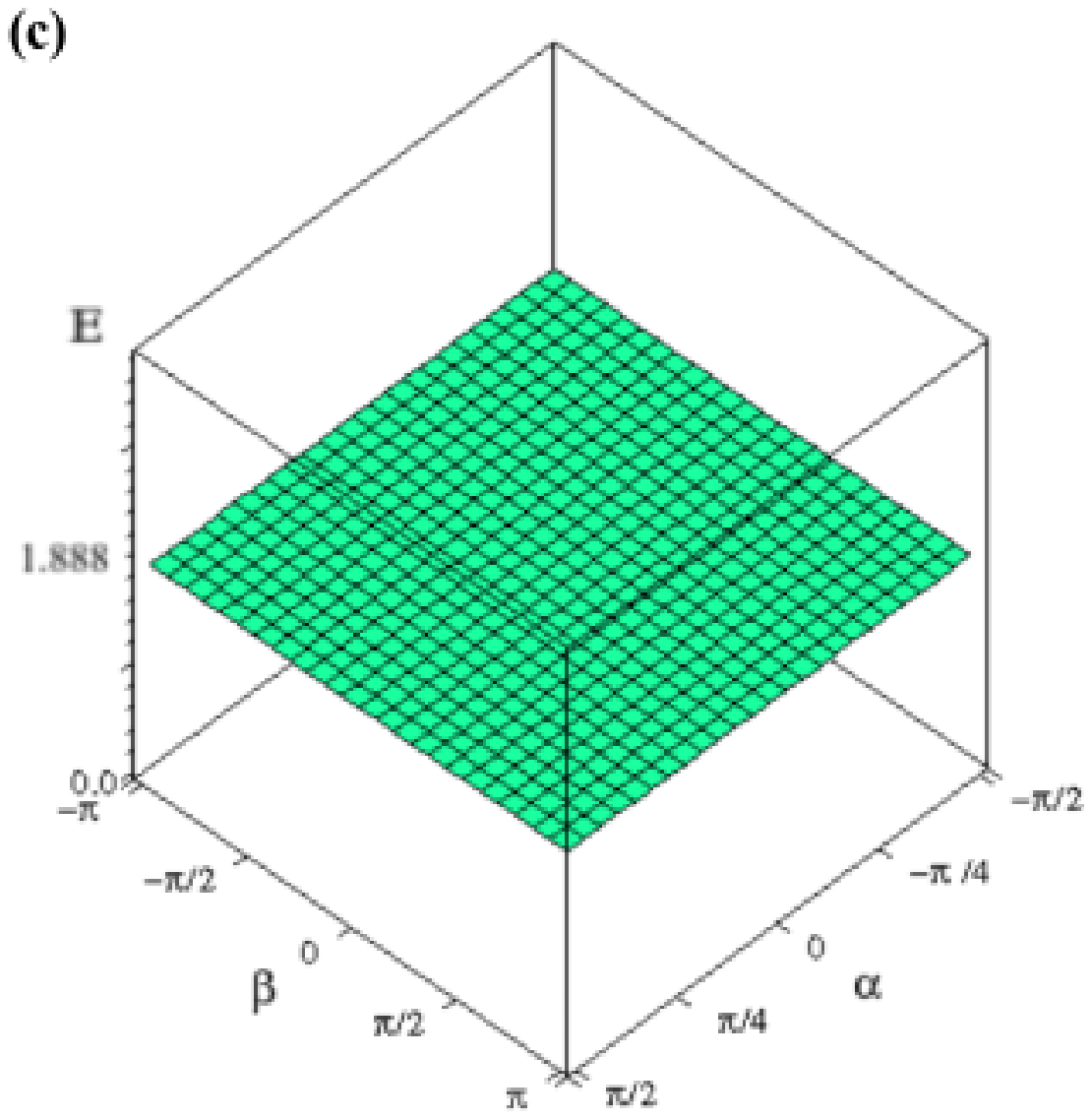}
\end{center}\vskip-5mm
\caption{\small Asymptotic entropy of entanglement $E$ for the non-local separable initial condition $\psi_s(\alpha,\beta)$, eq.~(\ref{eq:noloc1}),  with non-locality restricted to the $x$-direction. The surfaces correspond to different initial coin states: (a) $\ket{\Psi^+}$, (b) $\ket{LR}$ and (c) $\ket{\Psi^-}$, defined in eq.~(\ref{eq:bell1}). In each case, the flat surface indicates the CPE level for a localized initial position with the same initial coin state.}
\label{fig:non-local-separable}
\end{figure}

Up to this point, only localized initial conditions have been considered. In the one-dimensional case it is known that non-local initial states can modify significantly the dynamics of the QW. For instance, the Survival Probability (the probability of finding the walker in a given region which includes the starting point) decays as $t^{-1}$ for localized initial states and as  $t^{-3}$ for non-local initial states with appropriate relative phases \cite{Abal06}, so the probability flux spreads out faster than in the localized case. In the specific case of asymptotic CPE, previous work for a one-dimensional QW suggests that non-local initial states can give rise to a broader range of variation for CPE levels \cite{Abal06-ent}. However, this initial work was limited in scope because the one-dimensional case does not allow for initially entangled (coin or position) states as the 2D case does. We now consider initial sates of the form 
\begin{equation}\label{eq:noloc-istate}
 \ket{\Psi(0)}=\ket{\psi(\alpha,\beta)}\otimes\ket{\chi}
\end{equation} 
\begin{figure}
\begin{center}
\includegraphics[width=5.5cm]{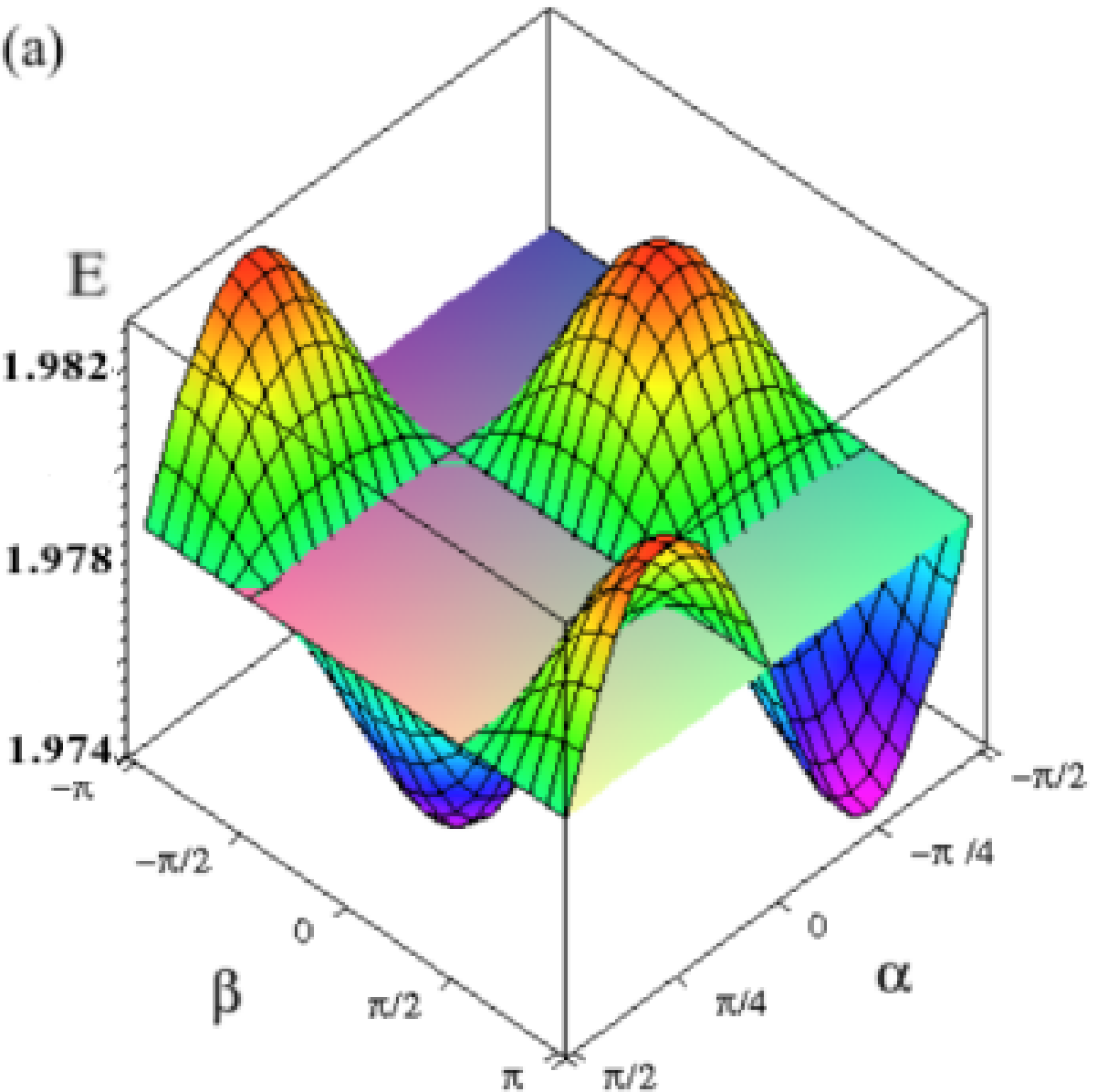}\\
\includegraphics[width=5.8cm]{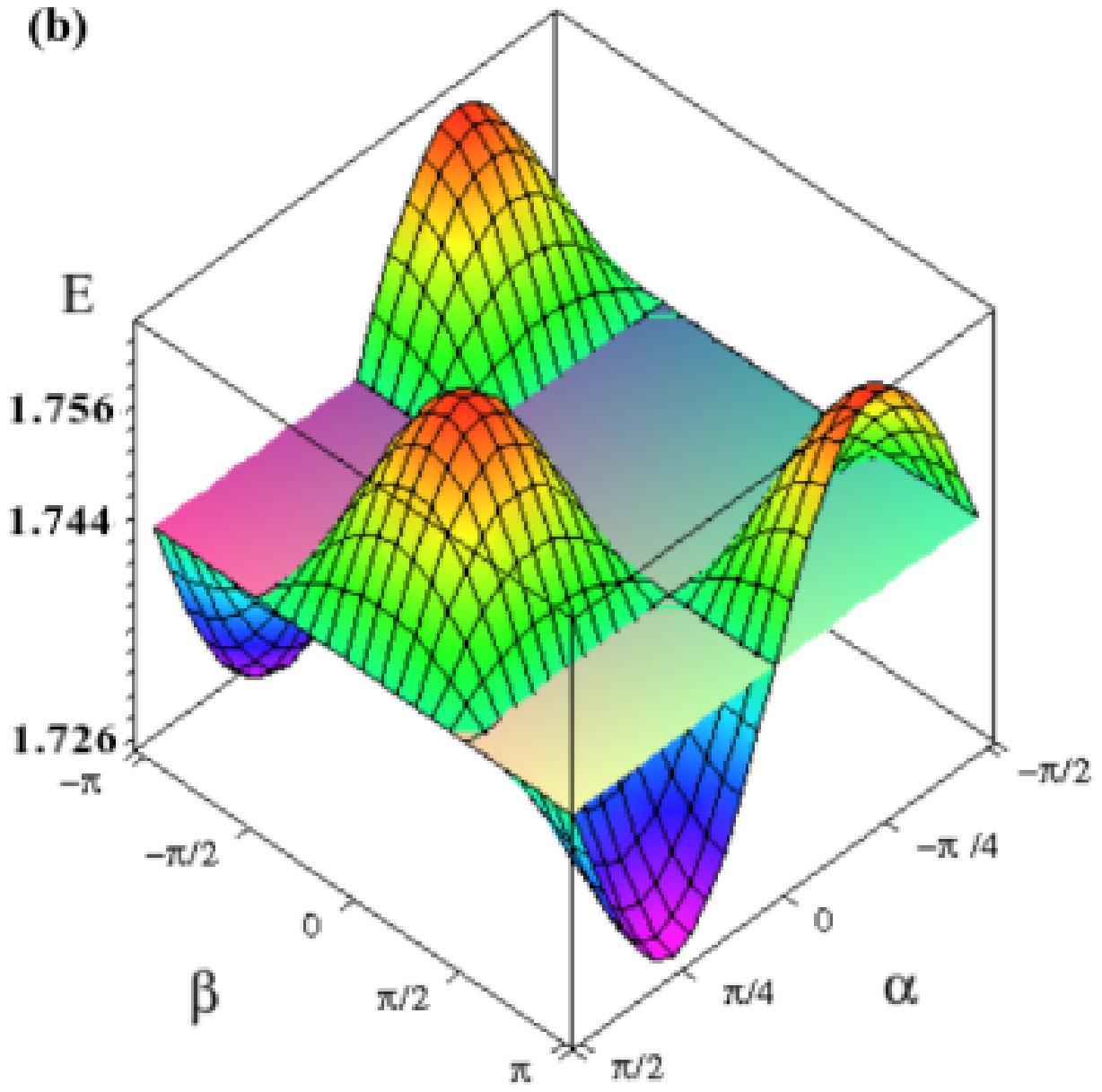}\\
\includegraphics[width=5.5cm]{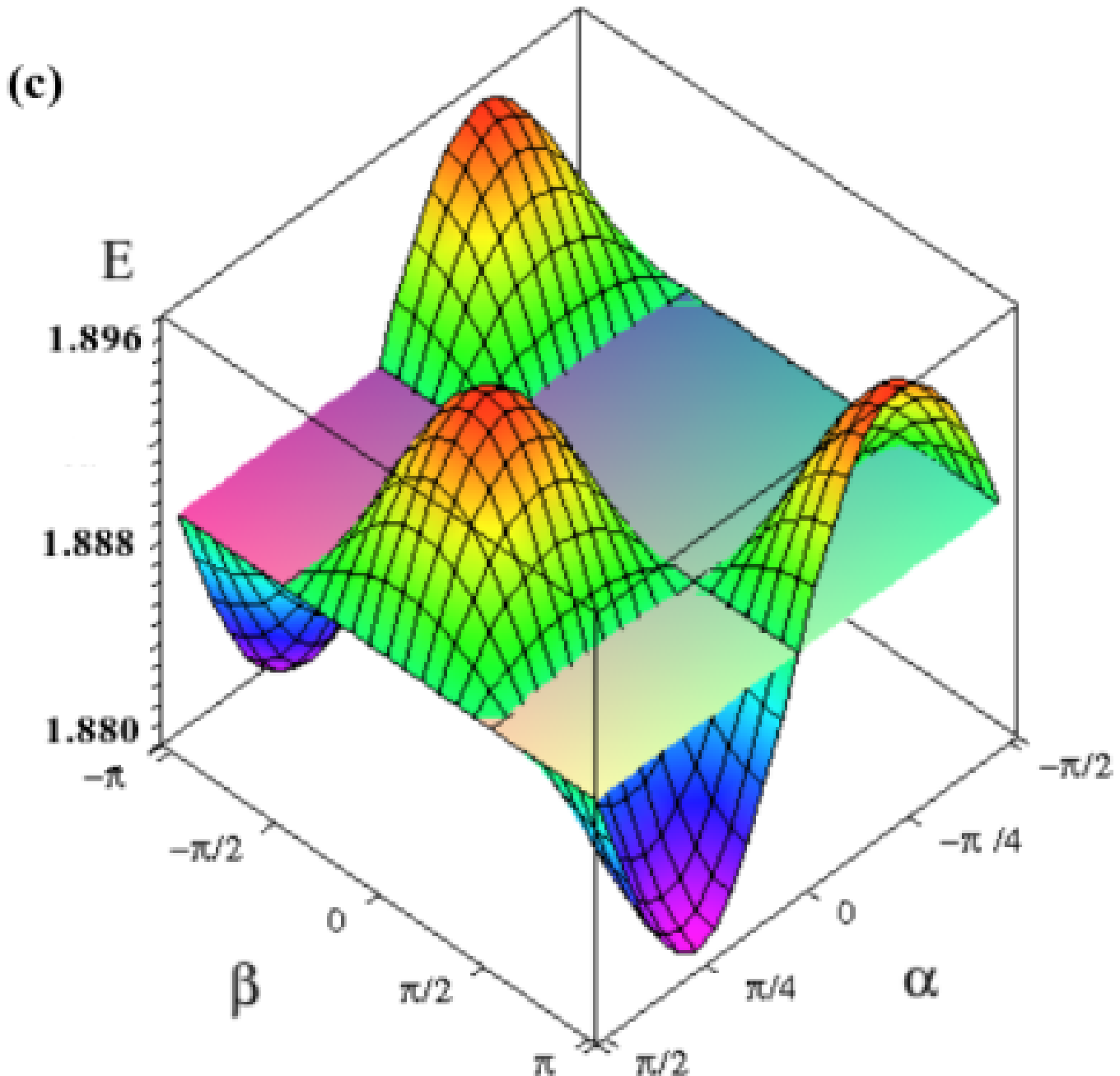}
\end{center}
\caption{\small Asymptotic entropy of entanglement $E$ for the non-local entangled initial condition $\psi_e(\alpha,\beta)$, eq.~(\ref{eq:noloc2}). The surfaces correspond to three different initial coin states as described in Fig.~\ref{fig:non-local-separable}. In each case, the flat surface indicates the CPE level for a localized initial position with the same initial coin state.}\label{fig:non-local-entangled}
\end{figure}
where the coin state $\ket{\chi}$ is chosen within the family $\ket{\chi^{(II)}}$ as either one of: the Bell states  $\ket{\chi}=\ket{\Psi^{\pm}}$ defined in eq.~(\ref{eq:bell1}) or the product state $\ket{LR}$. The initial position $\ket{\psi(\alpha,\beta)}$ is a superposition state on the two-dimensional lattice. We start by considering the effects of  small non-locality, with the initial states 
\begin{eqnarray}
 \ket{\psi_s}&\equiv&\left(\cos\alpha\ket{-1}_x +e^{i\beta}\sin\alpha\ket{1}_x\right)\otimes\ket{0}_y\label{eq:noloc1}\\
 \ket{\psi_e}&\equiv& \cos\alpha\,\ket{-1,1} + e^{i\beta}\sin\alpha\,\ket{1,-1}\label{eq:noloc2}
\end{eqnarray}
which describe non-local separable $(\psi_s)$ and entangled $(\psi_e)$ states in ${\cal H}_P$. In the later case, the degree of bipartite entanglement (PPE) is given by the entropy of entanglement, $S(\alpha)$, which can be read from Fig.~\ref{fig:compara-local}~(b) with the replacement $\theta\rightarrow\alpha$. In the next subsection we shall consider the effect of extended non-locality on CPE. 

For these non-local initial conditions the Fourier transformed amplitudes, $\tilde a(\mbf{k})=\sum_{\mbf{r}} e^{-i\mbf{k}\cdot\mbf{r}} a(\mbf{r})$, must be calculated. The relevant expressions are  
\begin{eqnarray}\label{eq:FT2}
 |\tilde a_s(\mbf{k})|^2&=&1+\sin(2\alpha)\cos(2k_x+\beta)\\
 |\tilde a_e(\mbf{k})|^2&=&1+\sin(2\alpha)\cos[2(k_x-k_y)+\beta].\nonumber
\end{eqnarray}
The calculation of the reduced density operator is simplified by noting that eqs.~(\ref{eq:Pij}) and (\ref{eq:F-nolocal}) imply 
\begin{equation}
 {\cal P}^\prime(\mbf{k})=|\tilde a(\mbf{k})|^2{\cal P}(\mbf{k}).
\end{equation} 
Thus, the matrix ${\cal P}^\prime(\mbf{k})$ for the non-local case is expressed in terms of the corresponding matrix ${\cal P}(\mbf{k})$ for the local case with the same initial coin state. The corresponding density operators are obtained from integration in $\mbf{k}$-space of the ${\cal P}(\mbf{k})$ previously calculated for the local case, with integrating factors given by eqs.~(\ref{eq:FT2}),
\begin{equation}\label{eq:hat_rho_C-noloc}
\hat\rho_c=\int\frac{d^2\mathbf{k}}{4\pi^2}\,|\tilde a(\mbf{k})|^2{\cal P}(\mbf{k}).
\end{equation}
Diagonalization of this operator leads to the corresponding asymptotic entropy of entanglement, $E (\chi;\alpha,\beta)$, which depends on the initial coin state and the two parameters that specify the initial position.

A comparison of the resulting CPE surfaces $E(\alpha,\beta)$ with the corresponding localized cases values of $E$ show, in most cases, variations due to position non-locality. The results for separable non-locality $(\psi_s)$ are shown in Fig.~\ref{fig:non-local-separable} and those for entangled non-locality, $(\psi_e)$, in Fig.~\ref{fig:non-local-entangled}. The flat surfaces represent the CPE of the localized cases with the same initial coin states. The largest variations due to non-locality are associated to balanced superpositions, $\alpha=\pm\pi/4$. Notice that the relative phase of the initial coin state can cause significant changes in CPE. The overall effect of this restricted non-locality on CPE is small, but this was to be expected since we are considering only small delocalizations about the origin. In the next section we consider the effect on CPE of non-local states with a large spread over the 2D lattice. 

\subsection{Extended non-locality}

We have considered initial position states with a small degree of non-locality, i.e. the initial  amplitudes that are non-zero only in a few sites around the origin and found that the changes in asymptotic CPE due to initial non-zero amplitudes at sites $x=\pm 1$ or $y=\pm 1$ as compared with the local case $x=y=0$, are small. However, is more extended non-locality capable of significant changes on the long-time CPE? Let us approach this question, considering an isotropic Gaussian position distribution for the amplitudes, $a(\mbf{r})\propto e^{-(x^2+y^2)/2\sigma^2}$, where $\sigma>0$ is a characteristic width. The Fourier-transformed amplitudes are $\tilde a(\mbf{k})=C e^{-\frac12 (k_x^2+k_y^2)\sigma^2}$ with  $C$ a constant obtained from the normalization condition $\protect{\int \frac{d^2\mbf{k}}{4\pi} \,|\tilde a(\mbf{k})|^2 = 1}$. Since
$$
\lim_{\sigma\rightarrow \infty} e^{-\frac12 (k_x^2+k_y^2)\sigma^2} = \frac{\pi}{\sigma^2}\,\delta (k_x)\,\delta (k_y)
$$
where $\delta(\cdot)$ is a Dirac's delta function, extended non-locality in position transforms the integration factor $|\tilde a(\mbf{k})|^2$ in a product of two delta functions and, in this limit, the integration in eq.~(\ref{eq:hat_rho_C-noloc}) becomes trivial. The resulting reduced density operator is
\begin{equation}
 \hat\rho_c(i,j)=\left[\sum_{\omega_k}\left\vert F(\omega_k)\right\vert^2 \alpha_i(\omega_k)\alpha^*_j(\omega_k)\right]_{k_x=k_y=0}.
\end{equation}
\begin{figure}
\hskip10mm\includegraphics[width=8.5cm]{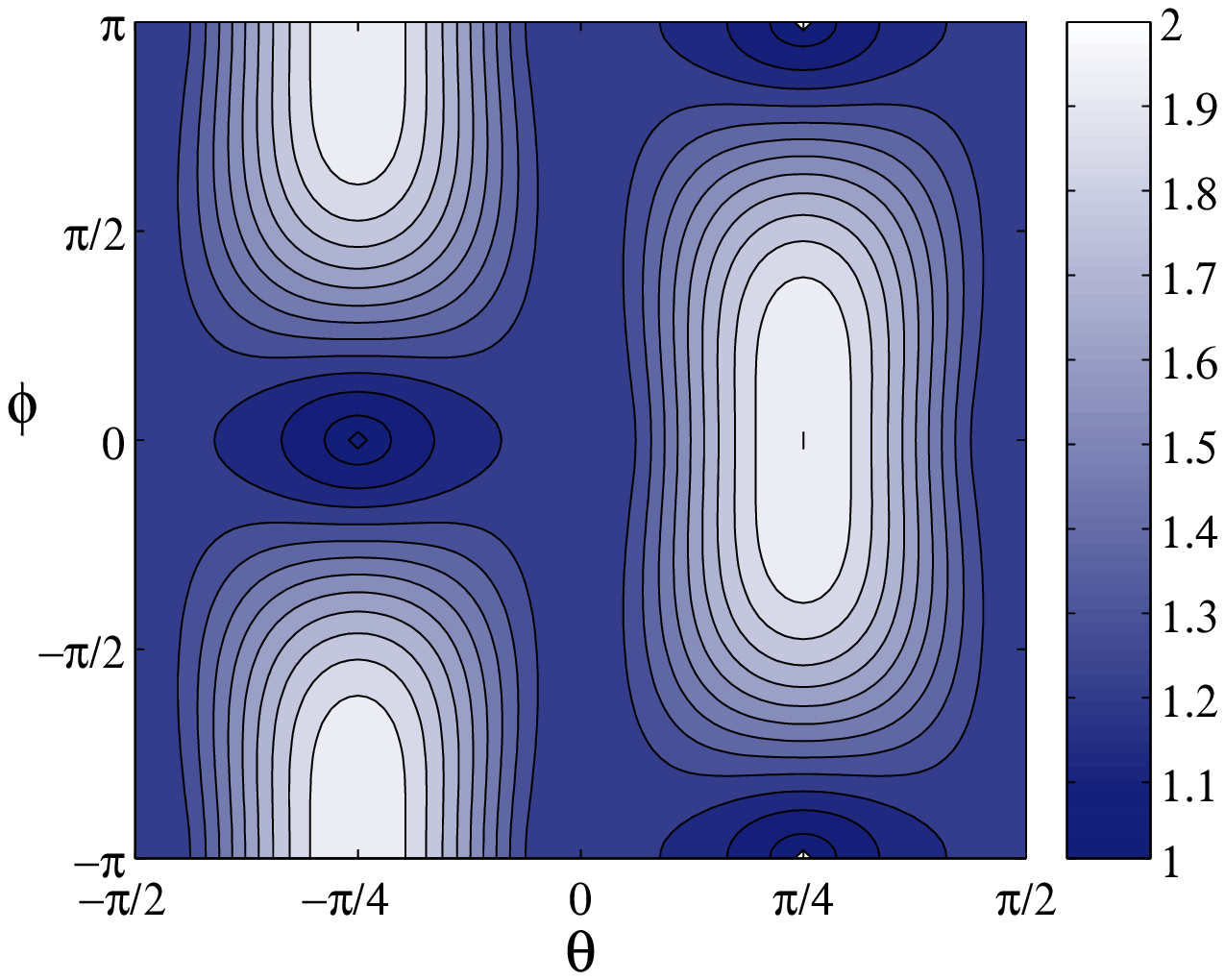}\\
\hskip-5mm\includegraphics[width=7cm]{./fig/gaussian-plot.eps}
\vskip-5mm
\caption{\small Upper panel: contour plot for $E(\theta,\phi)$ for initial coin $\chi^{(II)}(\theta,\phi)$ defined in eq.~(\ref{II}) and initial position uniformly distributed over the plane. Lower panel: constant $\phi$ sections for the same surface.}\label{fig:gaussian} 
\end{figure}

Thus for the extended position distribution, the asymptotic CPE depends only on the initial coin state which determines $|F|^2$. Let us choose the initial coin state within the family of entangled states $\chi^{(II)}(\theta,\phi)$, defined by eq.~(\ref{II}). In this case, the elements of the long-time reduced density operator, $\hat\rho_c$, result from evaluating the expressions for ${\cal P}(\mbf{k})$ obtained in the localized case, for $k_x=k_y=0$. The result can be expressed in simple form. The diagonal elements are
\begin{eqnarray*}
\hat\rho_c(1,1)&=&(\sin(2\theta)\cos\phi+3)/16\\
\hat\rho_c(2,2)&=&(1-\sin(2\theta)\cos\phi+8\cos^2\theta)/16\\
\hat\rho_c(3,3)&=&(9-\sin(2\theta)\cos\phi-8\cos^2\theta)/16\\
\hat\rho_c(4,4)&=&(3+\sin(2\theta)\cos\phi)/16,
\end{eqnarray*}
with $\tr{}{\rho_c}=1$, as expected. The expressions for the off-diagonal elements are,
\begin{eqnarray*}
 \hat\rho_c(1,2)&=&  (1-4\cos^2\theta+\sin(2\theta)\cos\phi)/16 \\
\hat\rho_c(1,3)&=& (-3+\sin(2\theta)\cos\phi + 4\cos^2\theta)/16 \\
\hat\rho_c(1,4)&=&  (-1+\sin(2\theta)\cos\phi)/16\\
\hat\rho_c(2,3)&=&(-1+\sin(2\theta)\cos\phi)/16 \\
\hat\rho_c(2,4)&=&(-1-\sin(2\theta)\cos\phi +4\cos^2\theta)/16 \\
\hat\rho_c(3,4)&=& (3-\sin(2\theta)\cos\phi-4\cos^2\theta)/16.
\end{eqnarray*}
The eigenvalues of this operator can be obtained analytically and the entropy of entanglement calculated as a function of the initial coin state. The resulting surface, $E(\theta,\phi)$, is shown in the upper panel of Fig.~\ref{fig:gaussian}. Two aspects of this surface are remarkable: (i) full asymptotic entanglement $E_{max}=2$ results for the initial coin $\ket{\Psi^+}$. Note that this maximum is basically flat in the $\phi$ direction, so it is robust against small variations in the relative phase of the initial coin state. And (ii) the minimum entanglement is now (exactly) $E_{min}= 1$ for the initial coin $\ket{\Psi^-}$, as shown in the lower panel of Fig.~\ref{fig:gaussian}. Finally, the initial product state $\ket{L,R}$ results in a low CPE level of $E\simeq 1.20$. Comparison with the local case in Fig.~\ref{FIG.L.PSI} shows that extended non-locality increases significantly the range of variation for asymptotic CPE. This result also contextualizes the small variations in CPE due to initial non-local positions in the neighborhood of the origin, as the variations may be larger when more extended initial states in position space are considered.

\section{Summary and Conclusions}
\label{sec:conclusions}

This work describes a method for the exact characterization of the long-time (asymptotic) coin-position entanglement (CPE) of a discrete-time quantum walk on a two-dimensional (2D) lattice or, alternatively, of two independent walkers on a line. In order to quantify the bipartite entanglement of the pure state $\rho$, the von Neumann entropy of the reduced density operator, $\rho_c$ is used. This quantity, $E(\rho_c)$, is scaled so that it varies between $0$ for a product coin-position state to $2$ for a fully entangled state. This corresponds to the $[0,1]$ variation of the one-dimensional case. The initial condition is allways chosen as a coin-position product state, so that no CPE is initially present. With this sole restriction, the general formalism leading to the exact calculation of the long time reduced density operator is presented. This treatment allows both local and non-local initial positions and arbitrary coin operations. The expressions we have presented can be readily applied to quantify the CPE of quantum walks in higher dimensions, such as the n-dimensional hypercube \cite{Moore,Marquezino}. The exact nature of our results allow us to distinguish between small variations in CPE and, for instance, clearly identify initial states which lead to full entanglement.

In order to illustrate the kind of results that can be obtained, we have considered in detail the case of a Hadamard coin operation, $U_C=H\otimes H$. Similar calculations can be done for any coin operation (i.e. Grover or DFT coins) for which the relevant eigenproblem in $\mathbf{k}$-space has been solved. 
We have first considered the case of localized positions with separable initial coin states. Then each walker starts in well defined states in the coin and position subspaces and the motion under a Hadamard coin remains separable. In this case, the CPE satisfies an additivity property which allows it to be expressed as the sum of the CPE for the corresponding one-dimensional motions. This additivity property can be generalized to the case of $N$ independent walkers with separable initial condition and coin operation. The CPE obtained is rather high, above 73\% of the maximum value in all cases considered, except for the extended non-locality discussed below. The maximum CPE depends on wether one can tune both initial coins or not. If both coins are tunable, the maximum CPE, $E=2$, can be obtained. 

When entangled initial coin or position states are considered, the motion is no longer separable and the calculation of CPE can not be reduced to the one-dimensional case. In this case, there is the (coin-coin or CCE) entanglement present in the initial state and the long-time CPE generated by the evolution. These are new results, as previous analytical work based on one-dimensional walks \cite{Abal06-ent} did not allow for this possibility. Extreme values for CPE (maxima or minima) appear associated with maximum values of CCE. Asymptotic CPE and initial CCE are related in the sense that maximum CPE implies a maximally entangled (CCE) initial coin state and minimum CPE implies an initial product state (i.e. no CCE). However, there are maximally entangled initial coins which do not lead to maximum CPE. 

We have also consider the effect of non-local initial conditions. We have shown that the calculation of the asymptotic density operator for this case can be reduced to that of the local case with the same initial coin state, provided an additional integration factor is included in the final step. Initial superpositions of two close sites can produce small changes in CPE (smaller than 1\%) with respect to the localized case. However, when extended non-local states are considered, large variations of CPE with respect to the local case may be obtained. We have investigated an initial state with Gaussian amplitude distribution about the origin. In the uniform limit, the variation for CPE is large. Values between $E=1$ and $E=2$ can be obtained depending on the initial coin state. 

Many studies of entanglement are restricted to the few-qubit case. The 2D quantum walk involves a two-qubit coin subspace and a position register which must have several qubits in order to accommodate the evolution during a significant number of steps. Thus, the entanglement in these systems is a complex issue, involving different kinds of bipartite entanglement (some of which we code-named as CCE, PPE or CPE). The analytical results presented here are a step towards a better understanding of the dynamics of entanglement in quantum walks. The relation between asymptotic CPE and the initial state is important, for instance, in order to gain insight on how a partial measurement will affect the walk. It also allows one to choose the initial conditions that will lead to the desired entanglement level. Entanglement is a key resource for quantum information processing, so we hope that these and similar results may stimulate new algorithmic applications.

\acknowledgements{We thank  R. Donangelo, M. Amini and M. Forets for useful discussions. G.A. acknowledges financial support from PEDECIBA (Uruguay) and Agencia Nacional de Investigación (ANII-Uruguay).}

\bibliography{qw}
\bibliographystyle{h-physrev} 

\end{document}